\documentclass[preprint,12pt]{elsarticle}

\usepackage{amssymb}
\journal{accepted for publication in Physica A}

\begin{document}

\begin{frontmatter}

%% Title, authors and addresses

%% use the tnoteref command within \title for footnotes;
%% use the tnotetext command for the associated footnote;
%% use the fnref command within \author or \address for footnotes;
%% use the fntext command for the associated footnote;
%% use the corref command within \author for corresponding author footnotes;
%% use the cortext command for the associated footnote;
%% use the ead command for the email address,
%% and the form \ead[url] for the home page:
%%
%% \title{Title\tnoteref{label1}}
%% \tnotetext[label1]{}
%% \author{Name\corref{cor1}\fnref{label2}}
%% \ead{email address}
%% \ead[url]{home page}
%% \fntext[label2]{}
%% \cortext[cor1]{}
%% \address{Address\fnref{label3}}
%% \fntext[label3]{}

\title{Coupled effects of market impact and asymmetric sensitivity in financial markets}

%% use optional labels to link authors explicitly to addresses:
%% \author[label1,label2]{<author name>}
%% \address[label1]{<address>}
%% \address[label2]{<address>}

\author{Li-Xin Zhong$^a$}\ead{zlxxwj@163.com}
\author {Wen-Juan Xu$^a$}
\author {Fei Ren$^b$}
\author {Yong-Dong Shi$^{a,c}$}

\address[label1]{School of Finance, Zhejiang University of Finance and economics, Hangzhou, 310018, China}
\address[label3]{School of Business, East China University of Science and Technology, Shanghai, 200237, China}
\address[label2]{School of Finance and Research Center of Applied Finance, Dongbei University of Finance and Economics, Dalian, 116025, China}

\begin{abstract}
%% Text of abstract
By incorporating market impact and asymmetric sensitivity into the evolutionary minority game, we study the coevolutionary dynamics of stock prices and investment strategies in financial markets. Both the stock price movement and the investors' global behavior are found to be closely related to the phase region they fall into. Within the region where the market impact is small, investors' asymmetric response to gains and losses leads to the occurrence of herd behavior, when all the investors are prone to behave similarly in an extreme way and large price fluctuations occur. A linear relation between the standard deviation of stock price changes and the mean value of strategies is found. With full market impact, the investors tend to self-segregate into opposing groups and the introduction of asymmetric sensitivity leads to the disappearance of dominant strategies. Compared with the situations in the stock market with little market impact, the stock price fluctuations are suppressed and an efficient market occurs. Theoretical analyses indicate that the mechanism of phase transition from clustering to self-segregation in the present model is similar to that in the majority-minority game and the occurrence and disappearance of efficient markets are related to the competition between the trend-following and the trend-aversion forces. The clustering of the strategies in the present model results from the majority-wins effect and the wealth-driven mechanism makes the market become predictable.

\end{abstract}

\begin{keyword}
%% keywords here, in the form: keyword \sep keyword
econophysics \sep market impact \sep asymmetric sensitivity \sep price fluctuations
%% MSC codes here, in the form: \MSC code \sep code
%% or \MSC[2008] code \sep code (2000 is the default)

\end{keyword}

\end{frontmatter}

%%
%% Start line numbering here if you want
%%
% \linenumbers

%% main text
\section{Introduction}
\label{sec:introduction}
Since Alfred Marshall, exploring the functional form of market impact has been a laboring but valuable task for economists\cite{moro}. More recently, to elucidate the discrepancy between the stock price shaped by excess demand and the actual transaction price, econophysicists have also got involved in the research of the roles of market impact in the evolution of stock prices\cite{stanley,mantegna,challet1}. Their work has opened a new window for the study of complex behaviors in  financial markets\cite{zhou,gronlund,yeung,podobnik,qiu,savit,toth}.

The market impact reflected in the increase and decrease of stock prices only tells us whether an excess demand exists or not, but not how it comes. Over the last decade, inspired by the findings in psychology that negative information should weigh more heavily on the brain than positive information, the roles of asymmetric sensitivity in the stock price performance have been studied by scientists\cite{veronesi,zhou2,plerou,mu}. It has been found that the investors often exhibit asymmetric responses to positive (gain) and negative (loss) information\cite{ren1,soroka}. They are prone to overreact to bad news and underreact to good news\cite{alwathainani,baker}. Such an effect may result in the change of the excess demand in the stock market.

To have a deep understanding of the evolutionary dynamics in financial markets, some agent-based models have been introduced in modeling the strategic interactions between the investors\cite{zhong1,johnson1,biswas}. Among them, the minority game (MG) provides us a simple yet effective way to model the evolution of stock prices\cite{zhang,challet2,xie,chen}. In the MG, the evolutionary mechanism is determined by two main factors: the global information and the individual strategy. At each time step, each agent makes a buying or a selling decision depending upon the historical price information and his own trading strategy. After all the agents have made their decisions, the stock price is updated according to the excess demand. To get more benefits in the investment, an agent will learn from his past mistakes and choose the best-performing strategy from his strategy pool as his decision-making strategy\cite{zheng, galla,ghosh,dhar}. Similar to the crowd-anticrowd problem in the MG, the herd behavior has also been studied in another repeated game, known as the Kolkata Paise Restaurant (KPR) problem\cite{ghosh2,ghosh3,chakrabarti}. Different from the two-choice competition in the MG, a macroscopic size of choices is considered in the KPR.

In the original MG, the individual strategy is a discontinuous variable. In exploring for the evolutionary mechanism of the strategies, it is somewhat difficult to give the descriptive nature of the strategies depending upon such a variable. As an extension of the original MG, the evolutionary minority game (EMG) introduced by Johnson et al has incorporated continuous strategy sets into the MG. With such a continuous variable, the descriptive nature of the strategies is easy to be reflected, i.e. by the distribution or the standard deviation of the individual strategies\cite{johnson,zhong2,johnson2,hart1,hart2}. In the EMG, the individual strategy is represented by a probability $g\in [0,1]$. In the decision making, an agent follows the outcome which can be predicted from the historical information with probability $g$ and does the opposite with probability $1-g$. An individual's strategy evolves according to its score. If the score of the strategy is below a threshold, it is modified within a certain range. The coevolutionary mechanism of the strategies and the stock prices in the EMG provides us more observable variables in the study of the market movement.

Although the roles of the market impact in the evolution of the stock prices have been widely discussed, both the coupled effect of the market impact and the asymmetric sensitivity and the coevolutionary mechanism of the individual strategies and the stock prices are still short of in-depth understandings. To address the coevolutionary mechanism of the individual strategies and the stock prices under different environmental conditions, in the present model, we incorporate the market impact and the asymmetric sensitivity into the EMG. The major findings of the present study are as follows.

(1) The change of the market impact parameter $\beta$ can effectively affect the distribution of individual strategies. There exists a critical point $\beta_{c}$, below which the population tend to become clustering and above which the population tend to self-segregate into opposing groups.

(2) Both the individual strategy and the stock price are closely related to an individual's asymmetric response to gains and loses. With little market impact, the introduction of the asymmetry sensitivity leads to the occurrence of a single dominant strategy and a large price fluctuation. A linear relation between the standard deviation of the stock price changes and the average value of the strategies is found. With full market impact, the asymmetry sensitivity only leads to the disappearance of the dominant strategies but not the stability of the stock prices.

(3) Theoretical analyses show that the occurrence of large price fluctuations is related to the  majority-winning effect while the suppression of the large price fluctuations is related to the minority-winning effect. Under the conditions where market impact is small, the market movement is predictable and the agents in the majority wins. Large price fluctuations are easy to occur in such a system. With full market impact, the market movement is unpredictable and the agents in the minority wins. The stock price is somewhat stable and an efficient market is easy to occur in such a system.

This paper is organized as follows. In Section 2, the evolutionary minority game with market impact and asymmetric sensitivity is introduced. In Section 3, the simulation results of the coevolution of individual strategies and stock prices is presented and the roles of the market impact and the asymmetric sensitivity are discussed. In Section 4, the mechanisms for the movement of the dominant strategies is analyzed theoretically. The conclusions and an outlook of future studies are given in Section 5.

\section{The model}
\label{sec:model}
We consider a model of $N$ agents repeatedly trading in the stock market. Each agent has a trading strategy, also called gene value $g$. At each time step, each agent makes a decision of buying (+1), selling (-1) or taking a holding position (0) according to the previous $m$ outcomes of price movement and his trading strategy. For example, we use the symbols of $\uparrow$ and $\downarrow$ as the rise and the fall of the stock prices respectively. For $m=3$, ($\uparrow\uparrow\uparrow$)$\downarrow$ represents the history in the memory, which means the price movement is down after three steps of rise. Faced with the global information $\uparrow\uparrow\uparrow$, the agent with strategy $g$ will make his decision following the prediction $\downarrow$ with probability $g$ and rejecting the prediction with probability $1-g$. After all the agents have made their decisions, the stock price is updated according to the equation

\begin{equation}
P(t+1)=P(t)+sgn[A(t)]\sqrt{\mid A(t)\mid},
\end{equation}
in which $A(t)=\sum_{i=1}^{N}a_i(t)$, $a_i(t)$ is the decision of agent $i$\cite{yeung,ren2}. At a given time, if there are more buyers than sellers, the stock price increases. If there are more sellers than buyers, the stock price decreases. The price information is stored in each agent's memory, which helps him make his prediction of the price movement in the next time steps.

In the real financial market, the excess demand has more or less effect on the transaction prices. In the present model, the market impact is reflected in the transaction price, which is defined as\cite{yeung}
\begin{equation}
P_{tr}(t)=(1-\beta)P(t)+\beta P(t+1),
\end{equation}
in which the variable $\beta$ ($0\leq \beta \leq1$) is used to measure the degree of the market impact. For $\beta=0$, the transaction price becomes $P_{tr}(t)=P(t)$,  which indicates that the transaction price contains no market impact and is determined by the immediate price. For $\beta=1$, the transaction price becomes $P_{tr}(t)=P(t+1)$, which indicates that the transaction price contains full market impact and is determined by the next price.

Each agent's strategy is modified according to the strategy score $d$, which is defined as an accumulated value of gains and losses after the strategy has been adopted.

\begin{equation}
d^+=\sum_{T^+=1}^{T_{max}^+}[P_{tr}(t_{sell})-P_{tr}(t_{buy})]_{T^+},
\end{equation}

\begin{equation}
d^-=\sum_{T^-=1}^{T_{max}^-}[P_{tr}(t_{sell})-P_{tr}(t_{buy})]_{T^-},
\end{equation}

\begin{equation}
d=d^++Rd^-,
\end{equation}
in which $T_{max}^+$ and $T_{max}^-$ are the transaction times corresponding to the cases of $P_{tr}(t_{sell})\geq P_{tr}(t_{buy})$ and $P_{tr}(t_{sell})<P_{tr}(t_{buy})$ respectively after the strategy has been adopted. $R (\geq 1)$ is the ratio demonstrating whether the asymmetric sensitivity exists or not. When $R=1$, the agents have the symmetric sensitivity to gains and loses.  When $R>1$, the agents have the asymmetric sensitivity and overreact to loses. There exists a predefined threshold $D$, if $d<D$, a new strategy is chosen from $[g-\varepsilon, g+\varepsilon]$ with an equal probability and the strategy score is reset to $d=0$. In the present model, the option to hold a position, that is, a decision of taking no buying or selling actions, is included in each agent's strategy $g$. For example, facing a global information that selling is beneficial, if an agent has a stock in his hand, he will sell it with probability $g$ and take no action with probability $1-g$. If the agent has no stock in his hand, he will buy the stock with probability $1-g$ and take no action with probability $g$. Therefore, between the time when an agent takes an action of buying and the time when he takes an action of selling, the agent holds a position. Such a mechanism is reflected in the equations (3-4) where the period of time $\mid t_{sell}-t_{buy}\mid$ is closely related to the gains or losses in each attempt.

In the evolution of individual strategies, the characteristics of the strategy distribution $P(g)$ can be represented by the standard deviation of the strategies, which satisfies the equation

\begin{equation}
\sigma_g=\sqrt{<g ^2>-<g>^2}.
\end{equation}
A small $\sigma_g$ implies that the agents tend to adopt the same strategy and the population cluster around a specific strategy. The larger the value of $\sigma_g$, the more dispersed the strategies.

The characteristics of the evolution of the stock prices can be represented by the price fluctuations, which can also be reflected in the standard deviation $\sigma_{P}$ of the price changes within a period of time. $\sigma_{P}$ is defined as

\begin{equation}
\sigma_{P}=\sqrt{<\delta P ^2>-<\delta P>^2}.
\end{equation}
A large $\sigma_{P}$ implies that the stock price is unstable and a large price fluctuation occurs. In such conditions, an agent's gains and loses depend upon whether he can make an accurate prediction of the market movement or not. If he can follow the movement of the market, he will attain more. If not, he will lose more.

The predictability of the rise and the fall of the stock prices is quite important for the investors to get more benefits. Therefore, how to measure it becomes an important work. Following the work done in \cite{yeung}, the predictability of the stock prices is defined as

\begin{equation}
H=\sum_{\mu=0}^{2^m-1} \rho(\mu)<\delta P \mid \mu>^2,
\end{equation}
in which $\mu$ is the possible state of the system,  $\rho(\mu)$ is the probability of the occurrence of $\mu$ and $<\delta P \mid \mu>$ is the conditional probability of the average value of the price change. If the value of $H$ is small, it implies that the price change is unpredictable and the stock market is an efficient market. If the value of $H$ is large, it implies that the price change is predictable and the stock market is an inefficient market.

Another parameter corresponding to whether an agent can make an accurate prediction of the rise and the fall of the stock prices is the winning probability, which is defined as

\begin{equation}
P_W=\frac{\Delta{T_W}}{\Delta T},
\end{equation}
in which $\Delta T$ is the statistical time window and $\Delta{T_W}$ is the winning times within it. In real society, an accurate prediction of the rise and the fall of the stock prices may attract more people to participate actively in buying and selling, which may lead to the occurrence of bubbles in the stock market.

\section{Results and discussions}
\label{sec:results}

\begin{figure}
\includegraphics[width=6cm]{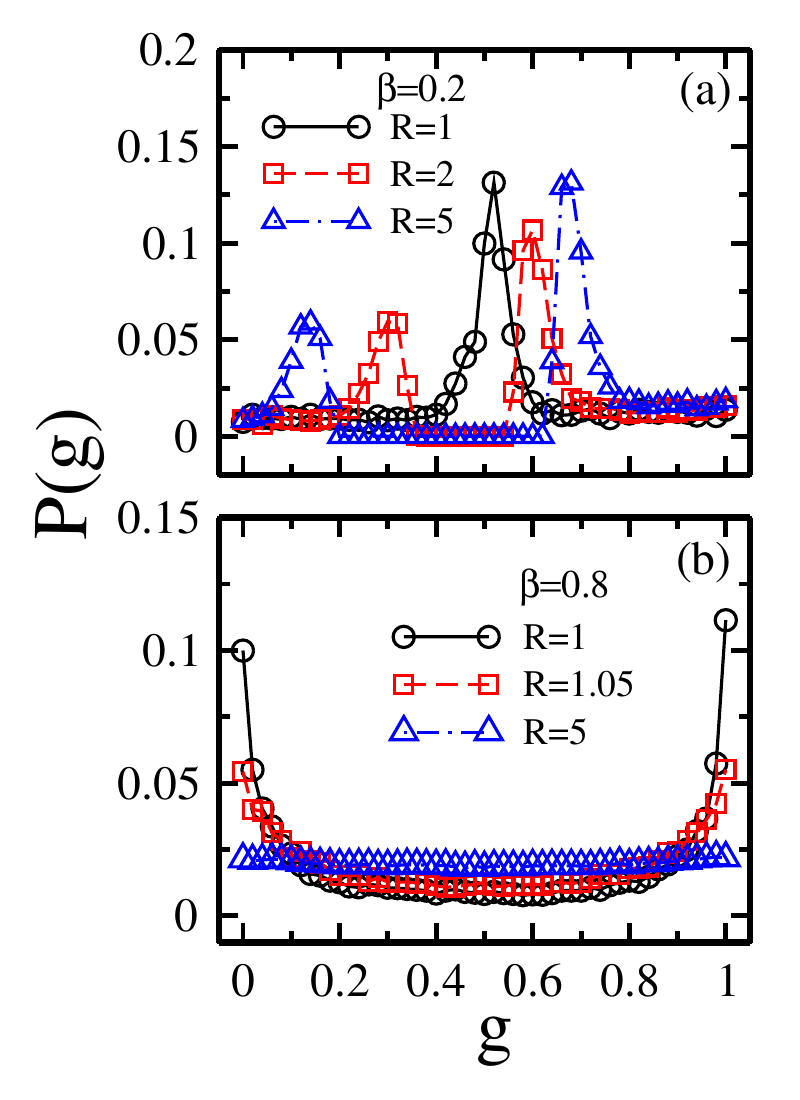}% Here is how to import EPS art
\caption{\label{fig:epsart}Strategy distribution P(g) with $N=101$, $m=3$, $D=-4$, (a) $\beta =0.2$, $R=1$ (circles), 2(squares), 5(triangles), and (b) $\beta =0.8$, $R =1$ (circles), 1.05 (squares), 5 (triangles). All the data are obtained by averaging over 100 runs and 1000 times after 100000 relaxation times in each run.}
\end{figure}

Figure 1 (a) and (b) show the long-time strategy distribution $P(g)$ for $\beta=0.2$, 0.8 and different $R$. It is observed that the characteristics of the strategies are closely related to the market impact parameter $\beta$ and the asymmetric sensitivity parameter $R$. For a small $\beta=0.2$ and $R=1$, the agents are prone to adopt the strategy $g\sim 0.5$. Increasing $R$ leads to the increase or decrease of the dominant strategy. As we have an eye on the unaveraged strategy distribution, we find that the $g< 0.5$ and $g> 0.5$ strategies can not coexist. Depending upon different initial conditions, there exist a critical point $g_c^>(>0.5)$ or $g_c^<(<0.5)$, only the strategies $g>g_c^>$ or $g<g_c^<$ are left in the final steady state for a specific run. Increasing $R$ leads to the increase of $g_c^>$ and the decrease of $g_c^<$. For a large $\beta=0.8$ and $R=1$, the agents are prone to self-segregate into opposing groups and a U-shape $P(g)$ distribution occurs. Increasing $R$ leads to the decrease of the extreme strategies of $g=0$ and $g=1$ and a uniform $P(g)$ distribution is observed for a big enough $R$.

\begin{figure}
\includegraphics[width=6cm]{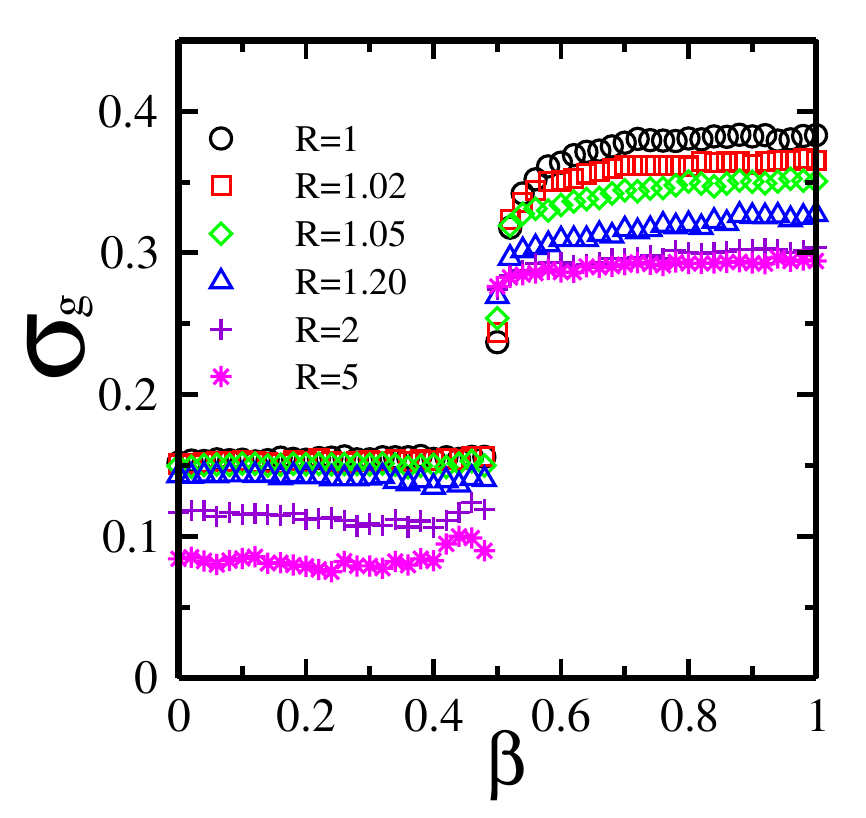}% Here is how to import EPS art
\caption{\label{fig:epsart}Standard deviation $\sigma_g$ of strategy distribution as a function of $\beta$ with $N=101$, $m=3$, $D=-4$ and $R =1$ (circles), 1.02 (squares), 1.05(diamonds), 1.2(triangles), 2(pluses), 5 (stars). All the data are obtained by averaging over 100 runs and 1000 times after 100000 relaxation times in each run.}
\end{figure}

Figure 2 displays the standard deviation $\sigma_g$ of the strategy distribution as a function of the market impact parameter $\beta$ for different $R$. In all the six cases, there exists a critical point $\beta_c\sim 0.5$. For $\beta <\beta_c$ and $R=1$, $\sigma_g$ keeps a small value of $\sigma_g\sim 0.15$. For $\beta >\beta_c$ and $R=1$, $\sigma_g$ keeps a large value of $\sigma_g\sim 0.38$. Increasing $R$ leads to the decrease of $\sigma_g$ within the range of $\beta <0.5$ and  $\beta >0.5$.

\begin{figure}
\includegraphics[width=8cm]{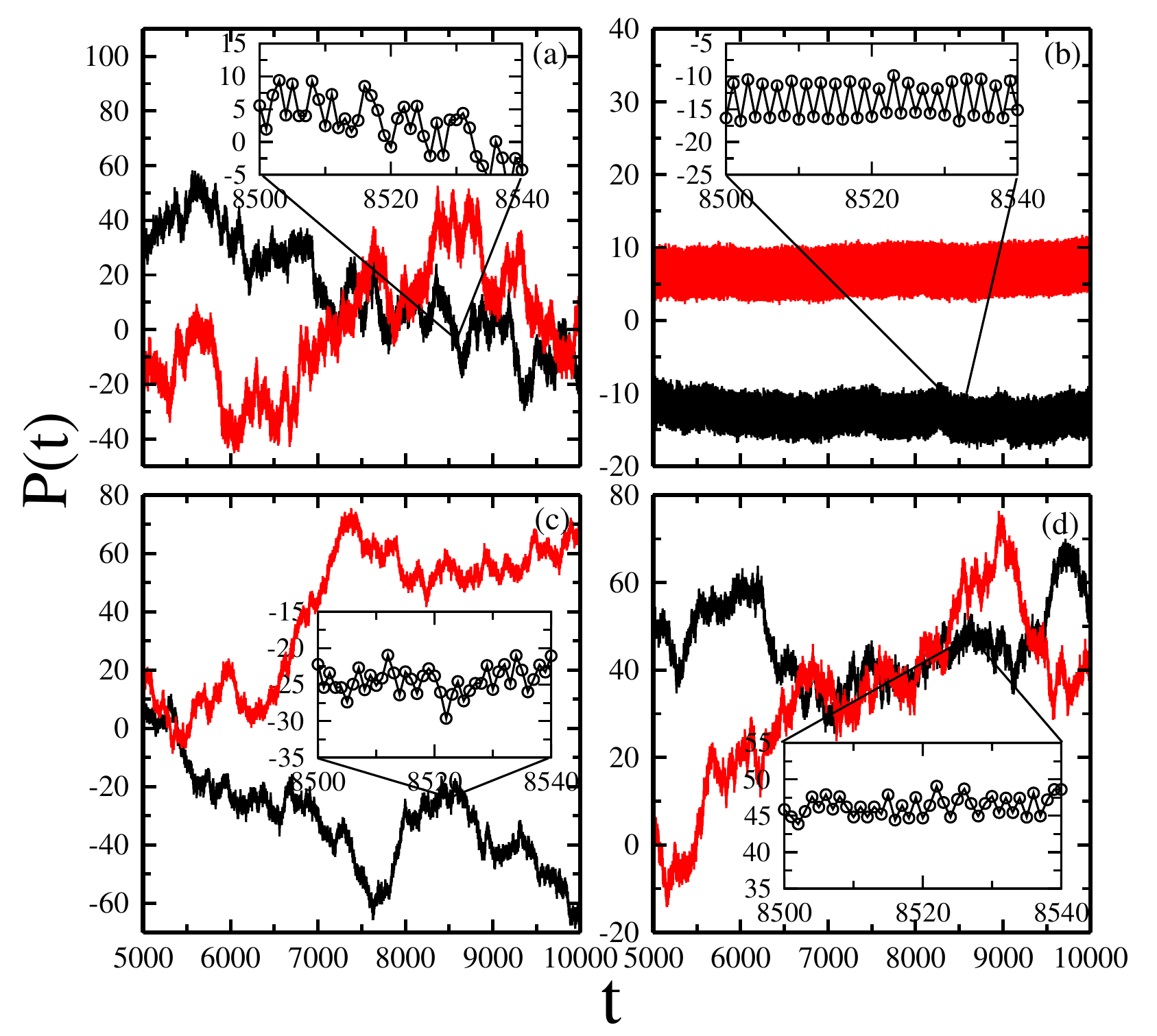}% Here is how to import EPS art
\caption{\label{fig:epsart}Evolution of the stock price $P(t)$ with $N=101$, $m=3$, $D=-4$, (a) $\beta$ =0.2,  $\overline p<0.5$, $R=1$(black), $R=5$(red); (b) $\beta$ =0.2,  $\overline p>0.5$, $R=1$(black), $R=5$(red); (c)$\beta$ =0.8, $R=1$(black), $R=1.05$(red); (d)$\beta$ =0.8, $R=1.2$(black), $R=5$(red).}
\end{figure}

Figure 3 shows the stock price $P(t)$ as a function of $t$ for $\beta=0.2$, 0.8 and different $R$. For a small value of market impact, i.e. $\beta=0.2$, which corresponds to the regime where the population cluster around a specific value of $g$ in fig.1(a). In the situations where all the agents adopt the strategies $g<g_c^<$ and the average value of the strategies satisfies $\overline g<0.5$, the stock prices exhibit large fluctuations. In the situations where all the agents adopt the strategies $g>g_c^>$ and the average value of the strategies $\overline g>0.5$, the stock prices exhibit zigzag oscillations. Increasing $R$ leads to the occurrence of a larger price fluctuation in both cases. For a large value of market impact, i.e. $\beta=0.8$, which corresponds to the regime where the strategy distribution changes from a U-shape to a uniform distribution in fig.1(b), the price fluctuations also exit. But compared with the price fluctuations in fig. 3(a) and (b), the oscillation amplitude becomes small. Incresing $R$ has no obvious effect on the change of the price fluctuations.

\begin{figure}
\includegraphics[width=6cm]{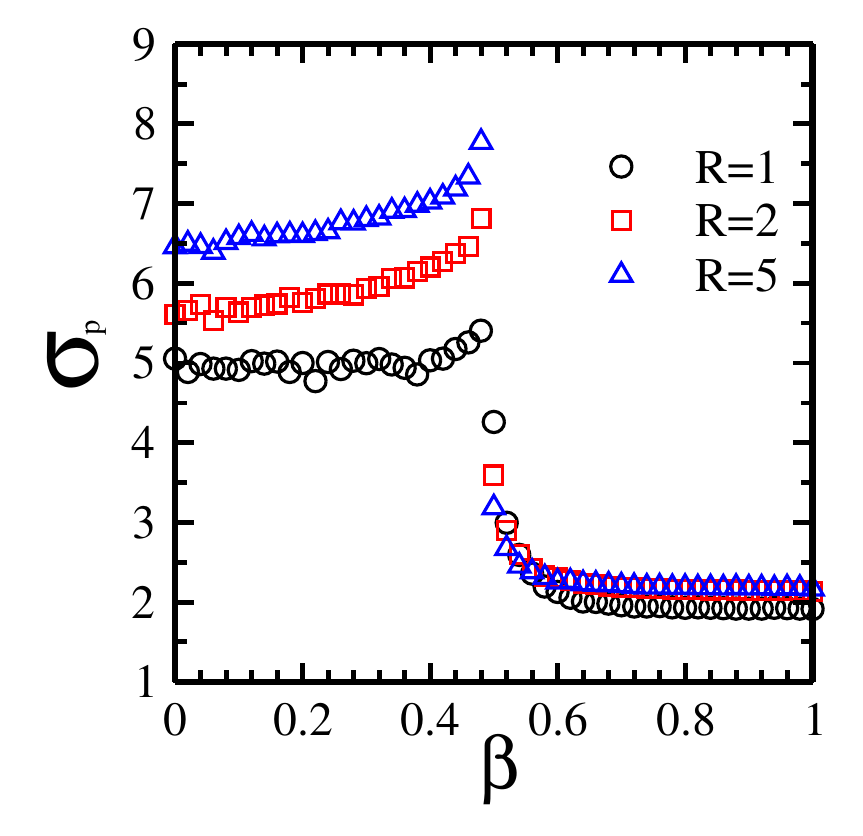}% Here is how to import EPS art
\caption{\label{fig:epsart}Standard deviation $\sigma_P$ of the stock price changes as a function of $\beta$ with $N=101$, $m=3$, $D=-4$ and $R=1$(circles), 2 (squares), 5 (triangles). All the data are obtained by averaging over 100 runs and 1000 times after 100000 relaxation times in each run.}
\end{figure}

In figure 4 we plot the standard deviation $\sigma_P$ of the stock price changes as a function of the market impact parameter $\beta$ for different $R$. Just as that in fig.2, for all the three cases of $R=1$, 2, and 5, there exists a critical point $\beta_c\sim 0.5$. For $R=1$ and $\beta_c< 0.5$, $\sigma_P$ keeps a large value of $\sigma_p\sim 5$.  For $R=1$ and $\beta_c> 0.5$, $\sigma_P$ keeps a small value of $\sigma_p\sim 1.8$. Increasing $R$ leads to an obvious increase of $\sigma_P$ within the range of $\beta<0.5$ and has no obvious effect on the change of $\sigma_P$ within the range of $\beta>0.5$. Comparing the results in fig. 2 with the results in fig.4, we find that both the individual strategies and the stock prices are affected by the market impact and the asymmetric sensitivity. Such results imply that there should be some close relation between the evolution of the stock prices and the evolution of the strategies in the present model.

\begin{figure}
\includegraphics[width=10cm]{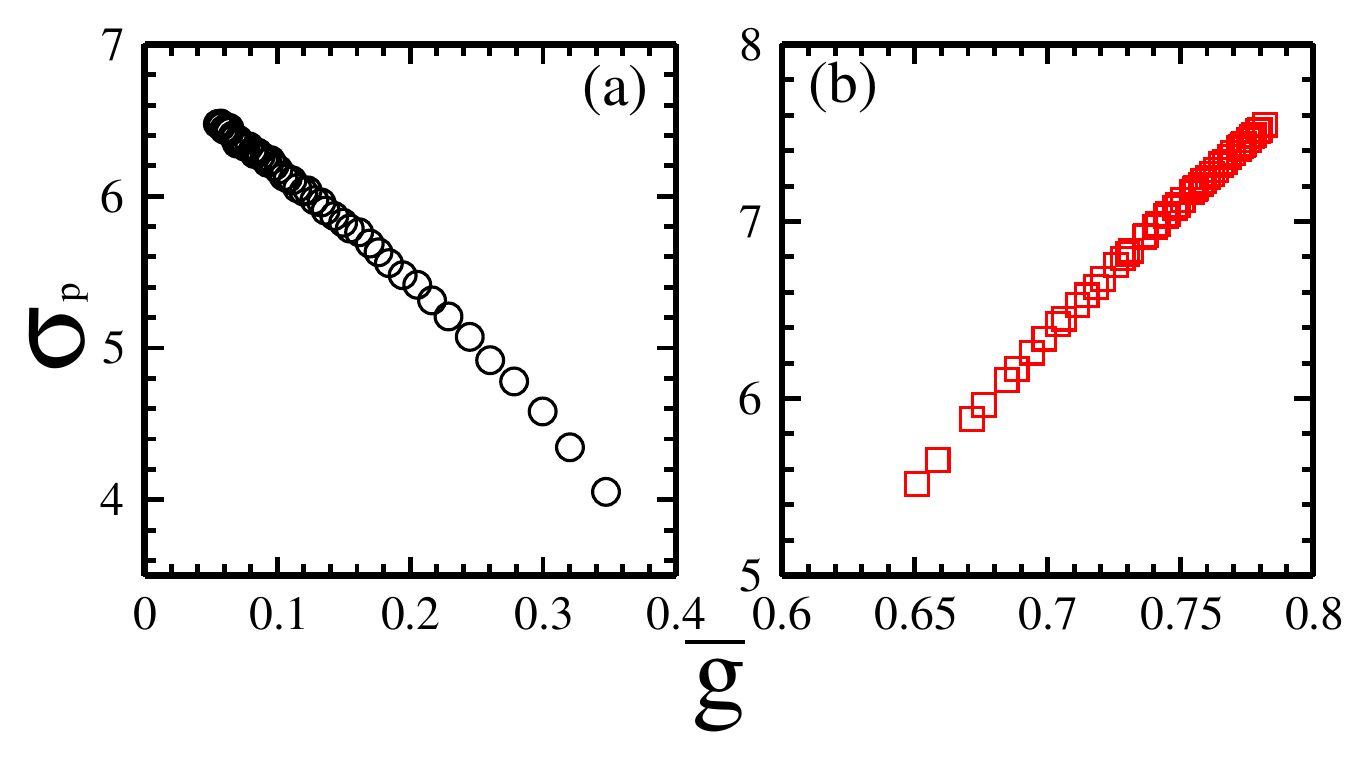}% Here is how to import EPS art
\caption{\label{fig:epsart}Standard deviation $\sigma_P$ of the stock price changes as a function of the average value of the individual strategies $\bar g$ with $N=101$, $m=3$, $D=-4$, $\beta =0.2$. (a)$\bar g<0.5$, (b) $\bar g>0.5$.}
\end{figure}

To find out whether there exists a functional relation between the change of the stock prices and the characteristics of the individual strategies, in fig.5 (a) and (b) we plot the standard deviation $\sigma_P$ as a function of $\bar g$ for $\beta =0.2$. The data in fig.5 (a) are obtained by averaging the runs in which all the strategies satisfy $g>g_c^>$ and the data in fig.5 (b) are obtained by averaging the runs in which all the strategies satisfy $g>g_c^>$. From fig.5 we observe that, in the conditions $\bar g<0.5$, $\sigma_P$ decreases linearly with the rise of $\bar g$. In the conditions $\bar g>0.5$, $\sigma_P$ increases linearly with the rise of $\bar g$. When we give fitted lines to the data in fig.5(a) and (b) respectively, we find they satisfy the equation $\sigma_p=a \bar p+b$, in which $a\sim 6.9489$, $b\sim -7.8047$ in fig.5(a) and $a\sim -4.2265$ and $b\sim15.093$ in fig.5(b). Such results indicate that, with little market impact, the evoltuion of the stock prices is closely related to the distribution of individual strategies. Given one of them we may accurately predict the other.

\begin{figure}
\includegraphics[width=6cm]{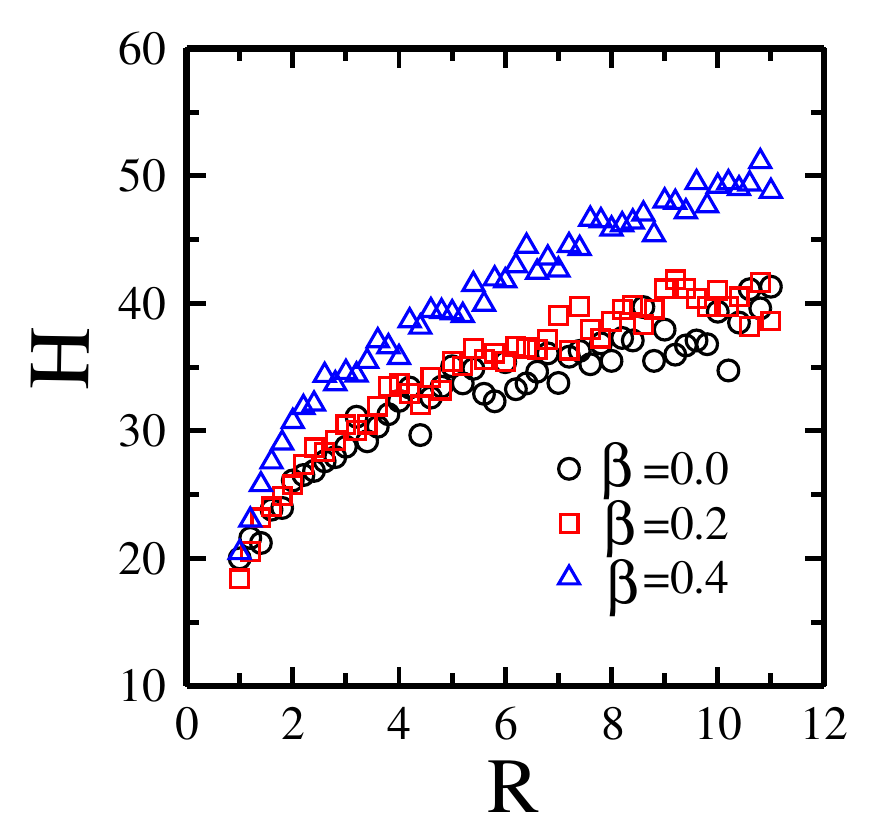}% Here is how to import EPS art
\caption{\label{fig:epsart}Predictability $H$ as a function of $R$ with $N=101$, $m=3$, $D=-4$ and $\beta$ =0(circles), 0.2 (squares), 0.4 (triangles). All the data are obtained by averaging over 100 runs and 1000 times after 100000 relaxation times in each run.}
\end{figure}

In fig.6 we give the predictability $H$ as a function of $R$ for different $\beta$. As $R$ increases from $R=0$ to $R=11$, for $\beta=0$,  $H$ increases from $H\sim 18$ to $H\sim 40$. Increasing $\beta$ has little effect on the change of $H$ for $R=1$ and leads to an obvious increase of $H$ for $R=11$. Such results indicate that the asymmetric sensitivity can effectively affect the movement of the market. In an inefficient market, the asymmetric responses to gains and loses will lead to the trend-following crowd effect and the evolution of the stock prices become more predictable.

\begin{figure}
\includegraphics[width=6cm]{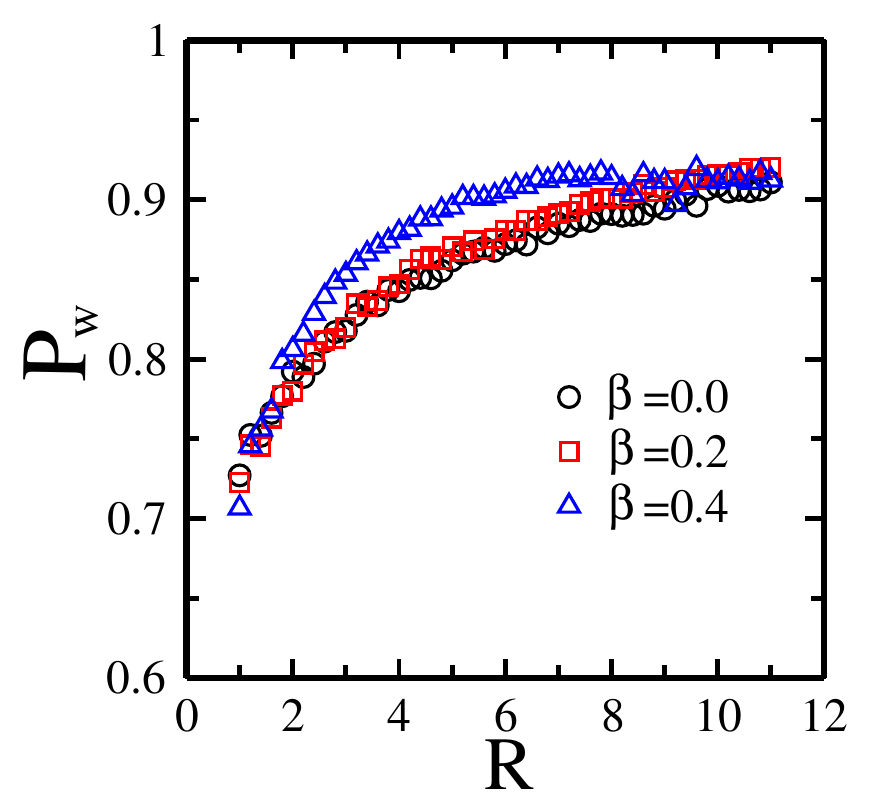}% Here is how to import EPS art
\caption{\label{fig:epsart}Winning probability $P_W$ (a) as a function of $R$ and (b) as a function of $H$ with $N=101$, $m=3$, $D=-4$ and $\beta$ =0(circles), 0.2 (squares), 0.4 (triangles). All the data are obtained by averaging over 100 runs and 1000 times after 100000 relaxation times in each run.}
\end{figure}

To find out whether the rise of predictability $H$ is beneficial for an agent to make an investment in the stock market, in fig.7 we plot the winning probability as a function of $R$ for different $\beta$. Just as that in fig.6, increasing $R$ can also effectively increase the winning probability. For $\beta=0$, as $R$ increases from $R=1$ to $R=11$, $P_W$ accordingly increases from $P_W=0.73$ to $P_W=0.91$. Increasing $\beta$ leads to an increase of $P_W$ for an intermediate $R$ but does not lead to an change of the maximum value of $P_W$. Such results indicate that, in a more predictable market, an investor is much easier to achieve success.

\section{Theoretical analysis}
\label{sec:analysis}
\subsection{\label{subsec:levelA}Phase transition from trend following to trend adverse}
In the present model, the evolution of the system exhibits quite different mechanisms below and above the critical point $\beta_c=0.5$. Below the critical point $\beta_c$, the system will evolve into the state where most of the agents adopt the same trading strategy and the stock price is unstable. But above the critical point $\beta_c$, the opposing groups are much easier to coexist and the stock price becomes stable. To theoretically understand the phase transition mechanism in the present model, we firstly make a comparison between the evolutionary mechanism in the minority game and the evolutionary mechanism in the majority game.

As individuals are engaged in the two-state game involving many other individuals, in the minority game, the agents in the minority group win. But in the majority game, the agents in the majority group win. Depending upon the evolutionary minority game, T.S.Lo et al have theoretically analyzed the relationship between the efficiency of the market and the distribution of individual strategies. The functional form of the strategy distribution has been found\cite{lo}, which satisfies

\begin{equation}
P(g)\varpropto \frac{1}{\frac{1}{2}-\tau(g)},
\end{equation}
in which $\tau(g)\sim \frac{1}{2}-\frac{1}{\sqrt N}g(1-g)$ is the winning probability of the agent with strategy $g$. From the above equation we find that the strategy distribution is closely related to the winning probability of different strategies. As $g$ changes from $g=0$ to $g=1$, $\tau(g)$ firstly decreases and then increases with the rise of $g$. For $g=0$ and $g=1$, $\tau(g)$ reaches its maximum value of $\tau(g)\sim \frac{1}{2}$. Therefore, $P(k)$ exhibits a U-shape distribution within the range of $g\in [0,1]$.

Different from that in the minority game, in the majority game, the agents following the crowd will win the game\cite{kozlowski,martino}. On condition that the agents do not interact, which corresponds to the situation where all the agents adopt the same strategy $g=\frac{1}{2}$ in the EMG, the difference between the number of agents in the majority group and the number of agents in the minority group should satisfy the relation $A\sim \sqrt N$\cite {kozlowski}. In the majority game, a macroscopic difference between the numbers of agents in different states should be $A=N$, which corresponds to the situation where all the agents make their decisions following or rejecting the historical price information in the EMG. Therefore, in the majority game, the strategy distribution should refrain from a symmetric distribution or clustering around $g=\frac{1}{2}$.

Then, we will give a comparison of the evolutionary processes below and above the critical point $\beta_c=0.5$. The evolutionary mechanism exists in the minority game and the evolutionary mechanism exists in the majority game are both found in the present model.

For the extreme case of $\beta=0$. From the transaction price equation, $P_{tr}(t)=(1-\beta)P(t)+\beta P(t+1)$, we can obtain $P_{tr}(t)=P(t)$, which indicates that, for a particular agent $i$, his winning probability should be determined by the instant price. Because each agent has known the instant price before he makes his decision, each agent should know of whether a buying or a selling decision in the next time step is beneficial for him or not. In such a case, the stock market becomes predictable and each agent can make good decisions according to the global information. Therefore, the majority-win situation found in the majority game will occur for $\beta=0$. For example, facing the global information $\uparrow\uparrow\uparrow$ and the history-dependent prediction $\uparrow\uparrow\uparrow\downarrow$, if most of the agents follow the prediction and sell the stocks, the stock price in the next time step decreases and the sellers win the game. If most of the agents reject the prediction and buy the stocks, the stock price in the next time step increases and the buyers win the game. Therefore, whether the agents make a decision of buying or selling, those in the majority win.

For the extreme case of $\beta=1$. From the transaction price equation we obtain $P_{tr}(t)=P(t+1)$,  which indicates that, for a particular agent $i$, his winning probability should be determined by the next price. Because each agent does not know the next price before he makes his dicision, each agent has no idea about whether a buying or a selling decision is beneficial for him or not. In such a case, it becomes difficult for each agent to make an accurate prediction of the market movement and the evolutionary mechanism is like that in the minority game. For example, facing the global information $\uparrow\uparrow\uparrow$ and the history-dependent prediction $\uparrow\uparrow\uparrow\downarrow$, if most of the agents follow the prediction and sell the stocks, the stock price in the next time-step decreases and the sellers lose the game. If most of the agents reject the prediction and buy the stocks, the stock price in the next time step increases and the buyers lose the game. Therefore, whether the agents make a decision of buying or selling, those in the minority win.

Within the range of $0<\beta<1$. Suppose the stock price in the time step $t$ is $P(t)$, if most of the agents buy the stocks in the next time step, the stock price in the time step $t+1$ should satisfy the inequality $P(t+1)\leq P(t)+\sqrt N$. In the time step $t+2$, because most of the agents bought the stocks in the last time step, the probability of price increase should be less than the probability of price decrease. Suppose most of the agents sell the stocks, the stock price in the time step $t+2$ should satisfy the inequality $P(t+2)\geq P(t+1)-\sqrt N$. Only considering the equality conditions, we find that the difference between the transaction prices in the sequential steps $t+1$ and $t+2$ satisfies the equation

\begin{equation}
P(t_{sell})-P(t_{buy})=\sqrt N(1-2\beta).
\end{equation}
The above equation shows that, within the range of $0<\beta<0.5$, the inequality $P(t_{sell})-P(t_{buy})>0$ is satisfied and the agents buying the stocks in the time step $t+1$ and selling the stocks in the time step $t+2$ win the game. Within the range of $0.5<\beta<1$, the inequality $P(t_{sell})-P(t_{buy})<0$ is satisfied and the agents buying the stocks in the time step $t+1$ and selling the stocks in the time step $t+2$ lose the game.

The above analysis shows that, within the range of $0\leq \beta<0.5$, the system is like a majority game and the agents in the majority group will win the game in the evolutionary process. Within the range of $0.5< \beta \leq 1$, the system is like a minority game and the agents in the minority group will win the game in the evolutionary process.

\subsection{\label{subsec:levelB}Relationship between the movement of the dominant strategy and the change of asymmetric sensitivity}
To get a theoretical understanding of how the fine-to-prize ratio $R$ affects the movement of the dominant strategy, we firstly divide the strategies into three groups: $g=0$, $g=\frac{1}{2}$ and $g=1$. For a large value of market impact $\beta\sim 1$, with which the system is like a minority game,  S.Hod et al have semi-analytically concluded that, as the fine-to-prize ratio $R$ increases, it is the temporal oscillation that results in the suppression of extreme strategies. In the present model, although not all the agents are active at a particular time, the evolution of the system is also determined by the difference between the number of agents buying the stocks and the number of agents selling the stocks, which is similar to that in the original EMG. Therefore, as $R$ increases, it should be the same mechanism that results in the occurrence of a uniform strategy distribution in the present model.

In the following, we mainly pay our attention on the small $\beta$ case and explore how the change of $R$ can lead to the occurrence of the extreme situations where most of the agents adopt the strategy of $g\sim 0$ or $g\sim 1$.

Firstly, let's give a comparison of the scores of different strategies, $g=0$, $g=\frac{1}{2}$ and $g=1$, in the evolutionary process. Considering the case of $\beta=0$. If all the agents adopt the strategy $g=\frac{1}{2}$, the number of the agents buying or selling the stocks is a random case and the excess demand should be proportional to $\sqrt N$. If all the agents adopt the extreme strategy $g=1$, they will make the same decision following the prediction, the excess demand should be proportional to $N$. If all the agents adopt the extreme strategy $g=0$, they will make the same decision rejecting the prediction, the excess demand should also be proportional to $N$. Because the scores of the strategies are determined by the excess demand, the above analysis indicates that both the score of strategy $g=0$ and the score of strategy $g=1$ should be larger than the score of strategy $g=\frac{1}{2}$. In the present model, because the evolution of the strategies is determined by the strategy score and the updating threshold, with the same updating threshold, the agents are more possible to adopt the strategies that are away from $g=\frac{1}{2}$.

Then, let's have a look at the changing tendency of the critical points $g_c^{<0.5}$ and $g_c^{>0.5}$ with a change of $R$ and $D$. As the system has evolved to the state where all the strategies satisfy $g>0.5$. For an agent $i$ with strategy $g_i$, he will win the game with probability $g_i$ and lose the game with probability $1-g_i$. The average score of the strategies should be proportional to $g_i-(1-g_i)R$. For a given population size $N$ and a threshold $D$, to refrain from being doomed in the evolutionary process, the score of strategy $g_i$ should satisfy a minimum value

\begin{equation}
g_i-(1-g_i)R\sim a(N)D,
\end{equation}
in which a(N) is determined by the population size and the time-dependent price equation. We obtain

\begin{equation}
g_i^{>0.5}\sim\frac{a(N)D+R}{1+R}.
\end{equation}
As the system has evolved to the state where all the strategies satisfy $g<0.5$, the agent with strategy $g_i$ will win the game with probability $1-g_i$ and lose the game with probability $g_i$, the score of strategy $g_i$ should satisfy a minimum value

\begin{equation}
(1-g_i)-g_iR\sim a(N)D.
\end{equation}
We obtain

\begin{equation}
g_i^{<0.5}\sim \frac{1-a(N)D}{1+R}.
\end{equation}

From the above two relations we find that the minimum value of $\mid g_i-0.5\mid$ is related to the asymmetric sensitivity parameter $R$ and the threshold $D$. Increasing $R$ and $D$ will lead to the increase of $g_i^{>0.5}$ and the decrease of $g_i^{<0.5}$. The above analysis also indicates that the higher the value of $\mid g_i-0.5\mid$, the higher the score of the strategy $g_i$. Therefore, with specific values of $R$ and $D$, the system will evolve to the state where nearly all the agents adopt the extreme strategy $g=0$ or $g=1$, which is in accordance with the simulation results.

\section{Summary}
\label{sec:summary}
The evolution of stock prices is related to complex correlations of various factors, including market movement and personal sentiment. The fundamental values of the stocks are easy to be found as the market moves to its stationary equilibrium, while the trend-following movement will drive the stock price far away from such an equilibrium. An extreme case is the occurrence of bubbles as a rush buying is found here and there.

In the attempt to get a deep understanding of the evolutionary dynamics of the prices in real financial markets, the minority game and its variants have been introduced to reflect the collective behaviors of trend followers or trend adverse. A variety of price patterns similar to that in the real markets have been found. However, in terms of the population movement, the original minority game does not give us a clear picture of the strategy structures. As a modification, the evolutionary minority game has employed an adjustable probability as the trading strategy and the population structures can be reflected in the strategy distribution.

By incorporating market impact and asymmetric sensitivity into the evolutionary minority game, we have examined the relation between the evolution of stock prices and the property of population structures under various complex market circumstances. The coupled effects of market impact and asymmetric sensitivity are reflected in the change of the relation between the fluctuations of stock prices and the distribution of individual strategies. With high market impact, an asymmetric response to gains and losses can effectively affect the strategy distribution but not the evolution of the stock prices. As the asymmetric sensitivity parameter $R$ increases, the strategy distribution changes from a U-shape distribution to a uniform distribution. But the stock price fluctuations have no obvious change with the rise of $R$. An efficient market, in which the prices are somewhat stable, occurs for high market impact. With low market impact, both the strategy distribution and the price fluctuations are affected by the asymmetric sensitivity. As $R$ increases, the system will finally evolve into the state where nearly all the agents adopt the extreme strategy $g=0$ or $g=1$ and the price fluctuations become large. A linear relation between the standard deviation of the stock price changes and the average value of individual strategies exists.

Theoretical analysis indicates that the evolutionary mechanisms below and above the critical point $\beta_c$ are determined by the majority-win effect and the minority-win effect respectively. Below the critical point $\beta_c$, the majority-win mechanism makes the evolution of the stock prices become predictable and the wealth-driven mechanism results in the movement of the dominant strategy. Above the critical point $\beta_c$, the minority-win mechanism leads to the occurrence of a crowd-anticrowd population structure and the occurrence of an efficient market.

By incorporating market impact and asymmetric sensitivity into the evolutionary minority game, the present model can effectively reflect the majority-minority effect in financial markets. However, the agents in the present model have to make their choices only regarding one stock, which is different from the situations in real financial markets where the agents can make their choices depending upon a variety of stocks. The KPR model, a variation of the MG model in which a macroscopic size of choices is considered as the agents take their actions\cite{ghosh,ghosh2,ghosh3}, may provide solutions to the issue. In the future, the asymmetric sensitivity and the heterogeneous communication structures will be further considered in the KPR and other evolutionary game models. Exploring the potentially successful strategies in different environmental conditions should be a special interest of ours.

\section*{Acknowledgments}
This work is the research fruits of the Humanities and Social Sciences Fund sponsored by Ministry of Education of China (Grant Nos. 10YJAZH137, 09YJCZH042), Natural Science Foundation of Zhejiang Province (Grant No. Y6110687), Social Science Foundation of Zhejiang Province ( Grant No.  10CGGL14YB) and National Natural Science Foundation of China (Grant Nos. 10905023, 11175079, 70871019, 71171036, 71072140).

%% The Appendices part is started with the command \appendix;
%% appendix sections are then done as normal sections
%% \appendix

%% \section{}
%% \label{}

%% References
%%
%% Following citation commands can be used in the body text:
%% Usage of \cite is as follows:
%%   \cite{key}          ==>>  [#]
%%   \cite[chap. 2]{key} ==>>  [#, chap. 2]
%%   \citet{key}         ==>>  Author [#]

%% References with bibTeX database:

\bibliographystyle{model1-num-names}
%\bibliography{<your-bib-database>}

\begin{thebibliography}{00}

\bibitem{moro}E. Moro, J. Vicente, L. G. Moyano, A. Gerig, J. D. Farmer, G. Vaglica, F. Lillo, and R. N. Mantegna, Market impact and trading profile of hidden orders in stock markets, Phys. Rev. E 80 (2009)066102.

\bibitem{stanley}H. E. Stanley, Book review: B. J. West and P. Grigolini, Complex webs: anticipating the improbable, Physics Today 64 (2011) 58-60 .

\bibitem{mantegna}R. N. Mantegna, H. E. Stanley, Scaling behaviour in the dynamics of an economic index, Nature 376 (1995) 46-49.

\bibitem{challet1}D. Challet, Inter-pattern speculation: beyond minority, majority and games, J. Econ. Dyn. Control 32 (2008) 85-100.

\bibitem{zhou}W.X. Zhou, The components of empirical multifractality in financial returns, Europhys. Lett. 88 (2009) 28004.

\bibitem{gronlund}A. Gronlund, I. G. Yi, B. J. Kim, Fractal profit landscape of the stock market, PLoS One 4(7) (2012) e33960.

\bibitem{yeung}C. H. Yeung, K. Y. Michael Wong, and Y.-C. Zhang, Models of financial markets with extensive participation incentives, Phys. Rev. E 77 (2008) 026107.

\bibitem{podobnik}B. Podobnik, H. E. Stanley, Detrended cross-correlation analysis: A new method for analyzing two nonstationary time series, Phys. Rev. Lett. 100 (2008) 084102.

\bibitem{qiu}T. Qiu, B. Zheng, F. Ren and S. Trimper, Return-volatility correlation in financial dynamics, Phys. Rev., E73 (2006) 065103.

\bibitem{savit}R. Savit, R. Manuca, and R. Riolo, Adaptive competition, market efficiency, and phase transitions, Phys. Rev. Lett. 82 (1999) 2203-2206.

\bibitem{toth}B. Toth, Z. Eisler, F. Lillo, J. Kockelkoren, J.-P. Bouchaud, J.D. Farmer, How does the market react to your order flow? Quantitative Finance 12(7) (2012) 1015-1024.

\bibitem{veronesi}P. Veronesi, Stock market overreaction to bad news in goods times: a rational expectations equilibrium model, The Review of Financial Studies 12(5) (1999) 975-1007.

\bibitem{zhou2}W.-X. Zhou, D. Sornette, Antibubble and prediction of China¡¯s stock market and real-estate, Physica A 337 (2004) 243¨C268.

\bibitem{plerou}V. Plerou, P. Gopikrishnan, X. Gabaix, H. E. Stanley, Quantifying stock-price response to demand fluctuations, Phys. Rev. E 66 (2002) 027104.

\bibitem{mu}G.-H. Mu, W.-X. Zhou, W. Chen, J. Kertesz, Order flow dynamics around extreme price changes on an emerging stock market, New J. Phys.
12 (2010) 075037.

\bibitem{ren1}F. Ren and L. X. Zhong, Price impact asymmetry of institutional trading in Chinese stock market,  Physica A 391 (2012) 2667.

\bibitem{soroka}S. N. Soroka, Good news and bad news: asymmetric responses to economic information, The Journal of Politics 68 (2) (2006) 372-385

\bibitem{alwathainani}A. M. Alwathainani, Does bad economic news play a greater role in shaping investors' expectations than good news? Global Economy and Finance Journal 3 (2)  (2010) 27 - 43.

\bibitem{baker}M. Baker, and J. Wurgler, Investor sentiment and the cross-section of stock returns, Journal of Finance 61 (2006) 1645-1680.

\bibitem{zhong1}L.X. Zhong, D.F. Zheng, B. Zheng, P.M. Hui, Effects of contrarians in the minority game, Phys. Rev. E 72 (2005) 026134.

\bibitem{johnson1}N.F. Johnson, S. Jarvis, R. Jonson, P. Cheung, Y.R. Kwong, and P.M. Hui, Volatility and agent adaptability in a self-organizing market, Physica A 256 (1998) 230.

\bibitem{biswas}S. Biswas, A. Ghosh, A. Chatterjee, T. Naskar, and B. K. Chakrabarti, Continuous transition of social efficiencies in the stochastic strategy minority game, Phys. Rev. E 85 (2012) 031104 .

\bibitem{zhang}Y.-C. Zhang, Modeling market mechanism with evolutionary games, Europhysics News 29 (1998) 51.

\bibitem {challet2}D. Challet, Y.-C. Zhang, Emergence of cooperation and organization in an evolutionary game, Physica A 246 (1997) 407 .

\bibitem {xie}Y. B. Xie, B. H. Wang, C. K. Hu, T. Zhou, Global optimization of minority game by intelligent agents, Eur. Phys. J. B 47 (2005) 587.

\bibitem {chen}K. Chen, B. H. Wang, and B.S. Yuan, Theory of the three-group evolutionary minority game, International Journal of Modern Physics B 18 (2004) 2387-2393.

\bibitem {zheng}D. F. Zheng, B. H. Wang, Statistical properties of the attendance time series in the minority game, Physica A 301 (2001) 560-566.

\bibitem{galla}T. Galla, and Y.-C. Zhang, Minority games, evolving capitals and replicator dynamics, J. Stat. Mech. (2009) P11012.

\bibitem{ghosh} A. Ghosh, D. De Martino, A. Chatterjee, M. Marsili, and B. K. Chakrabarti, Phase transitions in crowd dynamics of resource allocation, Phys. Rev. E 85 (2012) 021116.

\bibitem{dhar} D. Dhar, V. Sasidevan, and B. K. Chakrabarti, Emergent cooperation amongst competing agents in minority games, Physica A 390 (2011) 3477.

\bibitem{ghosh2} A. Ghosh, A. Chatterjee, M. Mitra, and B. K Chakrabarti, Statistics of the Kolkata Paise Restaurant problem, New Journal of Physics 12 (2010) 075033.

\bibitem{ghosh3}A. Ghosh and B. K. Chakrabarti, Kolkata Paise Restaurant (KPR) Problem (2009) [http://demonstrations.wolfram.com/KolkataPaiseRestaurant KPRProblem].

\bibitem{chakrabarti}A. S. Chakrabarti, B. K. Chakrabarti, A. Chatterjee, and M. Mitra, The Kolkata Paise Restaurant problem and resource utilization, Physica A 388 (2009) 2420-2426.

\bibitem{alfi}V. Alfi, A. De Martino, L. Pietronero, and A. Tedeschi, Detecting the traders¡¯ strategies in Minority-Majority games and real stock-prices, Physica A 382 (2007) 1-8.

\bibitem{johnson}N. F. Johnson, P. M. Hui, R. Jonson, and T. S. Lo, Self-Organized Segregation within an Evolving Population, Phys. Rev. Lett. 82(16) (1999) 3360-3363.

\bibitem{zhong2}L. X. Zhong, T. Qiu, B. H. Chen, C. F. Liu, Effects of dynamic response time in an evolving market, Physica A 388 (2009) 673-681.

\bibitem{johnson2}N. F. Johnson, M. Hart, and P. M. Hui,  Crowd effects and volatility in a competitive market. Physica A 269 (1999) 1.

\bibitem{hart1} M. L. Hart, P. Jefferies, P. M. Hui, and N. F. Johnson,  Crowd-anticrowd theory of multi-agent market games, European Physical Journal B 20 (2001) 547-550.

\bibitem{hart2} M. L. Hart, P. Jefferies, N. F. Johnson, and P. M. Hui, Generalized strategies in the minority game, Phys. Rev. E 63 (2000) 017102.

\bibitem{ren2}F. Ren and Y. C. Zhang, Trading model with pair pattern strategies, Physica A 387 (2008) 5523-5534.

\bibitem{lo}T. S. Lo, P. M. Hui, and N. F. Johnson, Theory of the evolutionary minority game, Phys. Rev. E 62(3) (2000) 4393-4396.

\bibitem{kozlowski}P. Kozlowski, and M. Marsili, Statistical mechanics of the majority game, J. Phys. A 36 (2003) 11725.

\bibitem{martino}A. De Martino, I. Giardina, and G. Mosetti, Statistical mechanics of the mixed majority-minority game with random external information, J. Phys. A 36 (2003) 8935.



%% \bibitem must have the following form:
%%   \bibitem{key}...
%%

% \bibitem{}

\end{thebibliography}

%% Authors are advised to submit their bibtex database files. They are
%% requested to list a bibtex style file in the manuscript if they do
%% not want to use model1-num-names.bst.

%% References without bibTeX database:

\end{document}